\def\PRL{{ Phys. Rev. Lett.\ }\/}
\def\PRB{{ Phys. Rev. B\ }\/}

\def\be{\begin {equation}}
\def\ee{\end {equation}}
\def\ber{\begin {eqnarray}}
\def\eer{\end {eqnarray}}
\def\bers{\begin {eqnarray*}}
\def\eers{\end {eqnarray*}}

\newcommand{\AAEDITOKAY}[2]{{}{\textcolor{black}{#2}}}

\makeatletter

\newcommand{\Rmnum}[1]{\expandafter\@slowromancap\romannumeral #1@}
\makeatother

\makeatletter
\newcommand*\env@matrix[1][*\c@MaxMatrixCols c]{%
  \hskip -\arraycolsep
  \let\@ifnextchar\new@ifnextchar
  \array{#1}}
\makeatother

\documentclass[aps,prb,showpacs,superscriptaddress,a4paper,twocolumn]{revtex4-1}

\usepackage{amsmath}
\usepackage{float}
\usepackage{color}

\usepackage[normalem]{ulem}
\usepackage{soul}
\usepackage{amssymb}

\usepackage{amsfonts}
\usepackage{amssymb}
\usepackage{graphicx}
\begin {document}

\title{Symmetry driven topological phases in XAgBi (X=Ba,Sr):  An \emph{Ab-initio} hybrid functional calculations}

\author{Chanchal K. Barman}
\thanks{These two authors have contributed equally to this work}
\affiliation{Department of Physics, Indian Institute of Technology, Bombay, Powai, Mumbai 400076, India}

\author{Chiranjit Mondal}
\thanks{These two authors have contributed equally to this work}
\affiliation{Discipline of Metallurgy Engineering and Materials Science, IIT Indore, Simrol, Indore 453552, India}

\author{Biswarup Pathak}
\email{biswarup@iiti.ac.in }
\affiliation{Discipline of Metallurgy Engineering and Materials Science, IIT Indore, Simrol, Indore 453552, India}
\affiliation{Discipline of Chemistry, School of Basic Sciences, IIT Indore, Simrol, Indore 453552, India}

\author{Aftab Alam}
\email{aftab@iitb.ac.in}
\affiliation{Department of Physics, Indian Institute of Technology, Bombay, Powai, Mumbai 400076, India}

\date{\today}

\begin{abstract}
Density functional theory (DFT) approaches have been ubiquitously used to predict topological order and non-trivial band crossings in real materials, like Dirac, Weyl semimetals and so on. However, use of less accurate exchange-correlation functional often yields false prediction of non-trivial band order leading to misguide the experimental judgment about such materials. Using relatively more accurate hybrid functional exchange-correlation, we explore a set of(already) experimentally synthesized materials (crystallizing in space group $P6_3/mmc$) Our calculations based on more accurate functional helps to correct various previous predictions for this material class. Based on point group symmetry analysis and {\it ab-initio} calculations, we systematically show how lattice symmetry breaking via alloy engineering manifests different fermionic behavior, namely Dirac, triple point and Weyl in a single material. Out of various compounds, XAgBi (X=Ba,Sr) turn out to be two ideal candidates, in which the topological nodal point lie very close to the Fermi level, within minimal/no extra Fermi pocket. We further studied the surface states and Fermi arc topology on the surface of Dirac, triple point and Weyl semimetallic phases of BaAgBi. We firmly believe that, while the crystal symmetry is essential to protect the band crossings, the use of accurate exchange correlation functional in any DFT calculation is an important necessity for correct prediction of band order which can be trusted and explored in future experiments.
 
\end{abstract}

\maketitle

\section{Introduction}

Discovery of topological semimetals/metals has revolutionalized the research field of condensed matter systems. This is mainly due to the fermionic excitations arising out of non-trivial band crossings which are protected by symmetries. Similar to gapless Dirac like bulk dispersion on the surface of topological insulators, the gapless dispersion of (semi)metals could even lead to host interesting surface-bulk correspondence by virtue of band touchings of non-degenerate conduction and valence bands.\cite{Murakami2007,discovery2014} Dirac, Weyl, Nodal line and triple point semimetals are examples of such topological semi(metals) featured from bulk band topology.\citep{Andrei2012,Alexey2015,Ashvin2018,CM2019,CKB2019,MoP2017,NLS-1,NLS-2,NLS-3,NLS-4,SCZHANG2017,RhSi2017,FeSi2018,dDirac2016} 
The band touching points in these semimetals are mediated by crystalline symmetries and thus they drop to stable topological character of the band crossings.\cite{dDirac2016,Bradlyn2016,TayRong2017,HaoZheng2017,Rappe2012} 

\begin{figure}[b]
\centering
\includegraphics[width=\linewidth]{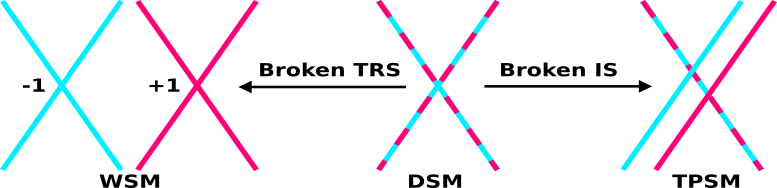}
\caption{Schematic diagram: a parent Dirac semimetal can transform into triple point semimetal  and Weyl semimetal due to the breaking of inversion (IS) and time reversal (TRS) symmetry.  }
\label{fig1}
\end{figure}

{\par}Till to date, over the past decade there has been active research  both from theory and experiment in distinct classes of topological semimetals, namely, Dirac semimetal (DSM),\cite{Ashvin2018,dDirac2016,Rappe2012} triple point semimetal (TPSM),\cite{TPSM2016,CM2019,CKB2019} and Weyl semimetal (WSM).\cite{Alexey2015,Ashvin2018,Bradlyn2016} However, achieving more than one type of topological semimetallic feature in a single compound are quite rare and still under active search. Since the band crossings in the topological semimetals are preserved via the unavoidable crystalline symmetries,\cite{dDirac2016,Bradlyn2016,TayRong2017,HaoZheng2017,Rappe2012} the idea of tuning a particular topological phase to another relies on gradual change of crystalline symmetries. For example, in Fig.~\ref{fig1}, a cartoon diagram is shown to describe how a Dirac semimetal gives birth to TPSM, and WSM under specific symmetry breaking. In a generic system with C$_{6v}$ point group symmetry along with both space inversion ($\mathcal{I}$) and time reversal symmetry (TRS), a pair of Kramer's pair linearly crosses to form a four fold degenerate Dirac node along the six-fold rotational (C$_6$) axis. The degeneracy of the Dirac nodes are enforced by the mirror ($\sigma_v$) and C$_6$ rotation, which will be described in details later in the manuscript. The breaking of $\mathcal{I}$ symmetry lifts the Kramer's degeneracy from a pair of band and form two triply degenerate nodal point under C$_{3v}$ little group. Further removal of TRS, transforms the Dirac node to a pair of Weyl nodes of opposite Chern number. 

{\par}Though the essential band crossings are solely determined by the material's space-group or point-group symmetry,\cite{FalsePositive} still the material predictions using Density-functional theory (DFT) approaches are quite challenging. This is because \AAEDITOKAY{the band crossings which we are after}{of the uncontrolled location of the band crossings which} may arise far away from the Fermi level and could be camouflaged within large density of states. More importantly, the use of inappropriate (less accurate) accurate exchange-correlation functional could lead to trivial gap between the band touching points. Thus, although the expected band crossings might be known from the space group of a material, these shortcomings still require a more refined search of new topological materials. Till to date, there has been a plethora of compounds predicted to be topological semimetals or insulators using conventional DFT approaches, many of them however had to be altered to be trivial insulator while using relatively more accurate exchange-correlation functional. For example, Zhang et al.\cite{Honeycomb2011} found topological non-trivial order in a series of compounds honeycomb structure, however, many of those compounds later proved to be trivial insulators.\cite{Honeycomb2013} J. Vidal et al. in Ref. \onlinecite{FalsePositive} showed that many compounds which were predicted to be topological insulators within DFT calculations using less accurate functional, are actually ``false positives" in reality. Hence, the conventional DFT predictions at the level of generalized gradient approximations (GGA) \AAEDITOKAY{could often leads to misguide the experimental search of materials that are non-trivial in topological order and could be the reason for the detriment of experiment-theory interaction.}{are misleading for future experimental search of topological materials.}

{\par} In this article, we choose a hexagonal compound BaAgBi which is already experimentally synthesized earlier and shown to crystallize in the space group P6$_3$/mmc (\# 194).\citep{synthesize2} Using the more accurate HSE06 hybrid functional calculation, we predicted this compound to be an ideal candidate for Dirac semimetal (DSM) with no extra Fermi pockets. Further, we discuss the interplay of symmetry in this class of materials, in general, and show how alloy induced symmetry breaking can lead to various low energy \AAEDITOKAY{excitonic}{exotic} phases such as topological Weyl, triple point semimetallic phases etc.  Apart from bulk, we have also simulated the surface states and the corresponding Fermi arcs for BaAgBi and it's alloy counterpart. In addition to BaAgBi which was a prototype for detailed analysis/calculations, we have also simulated a few other compounds belonging to the same class and also experimentally synthesized earlier,\citep{synthesize2,BaAgAsSynthesize} namely BaAgAs, SrCuBi, and SrAgBi. BaAgBi, SrAgBi and SrCuBi turn out to be the most promising candidates showing perfect DSM phase at ambient conditions, with no trivial Fermi pockets.

{\par} The importance of hybrid functional (HSE06) over conventional DFT calculations can be realized by closely analyzing the simulated band structure of BaAgAs.  Within the standard GGA calculation [\onlinecite{BaAgAs2019}], BaAgAs shows DSM behavior whereas our calculation, using HSE06 functional, predict it to be a trivial semiconductor with a band gap $\sim$0.30 eV (see Fig. S6 of SM\cite{supp}). CaAuAs is yet another example which shows a similar contrast between GGA and HSE06 calculation.\cite{supp,CaAuAs2018} As such, it is highly desired to perform hybrid functional calculation to correctly predict the topological non-trivial phase of any material and hence reliably guide the experimentalists for future measurements. For the purpose of comparisons, we have given GGA level band structures too in SM\cite{supp} for our prototype system.


\section{Computational Details}
First principle calculations were carried out using Density Functional Theory (DFT) implemented within the Vienna Ab Initio Simulation Package (VASP).\cite{GKRESSE1993,JOUBERT1999} Plane wave basis set using projector augmented wave (PAW)\cite{PEBLOCH1994} method was used with an energy cut-off of 500 eV. Generalized-gradient approximation by Perdew-Burke-Ernzerhof (PBE)\cite{JOUBERT1999} was employed to describe the exchange and correlation. Total energy (force) criterion was set upto 10$^{-6}$ eV (0.01 eV/\AA). The Brillouin zone (BZ) integration was performed using a 10$\times$10$\times$6 $\Gamma-$centered k-mesh. To accurately probe the band order and the band gap, we employed HSE06\cite{hse06} hybrid functional with 25\% of the exact exchange (mixing parameter AEXX=0.25, which is a default value). The relativistic spin orbit coupling (SOC) was included in all the calculations. Maximally localized Wannier functions (MLWFs)\cite{mlwf1,mlwf2,mlwf3} obtained from Wannier90 package,\cite{w90} were used to construct a tight-binding (TB) Hamiltonian. The topological properties including surface spectrum, Fermi arcs, and topological index were calculated using the iterative Green's function\cite{greenfn1,greenfn2,greenfn3} approach implemented in Wannier-Tools package.\cite{WTools} The optimized and experimental lattice parameters are given in SM.\cite{supp}

\section{Symmetry arguments} 

\begin{figure}[t!]
\centering
\includegraphics[width=0.7\linewidth]{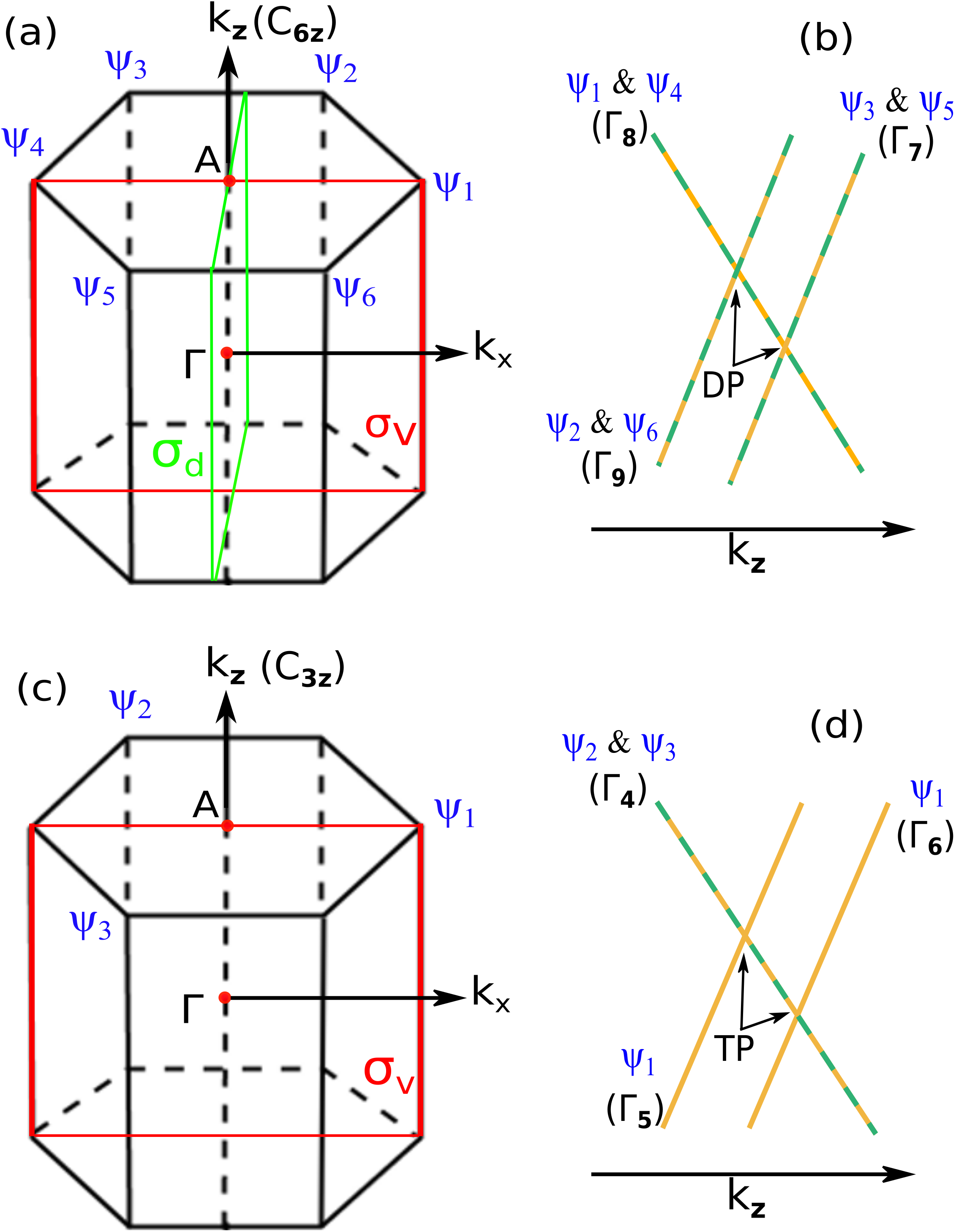}
\caption{(Color online) Schematic representation of high symmetry axis and mirror planes for (a) C$_{6v}$ and (c) C$_{3v}$. (b,d) Schematic representation of four-fold Dirac (DP) and three-fold triple point (TP) band crossings along $\Gamma$-A direction in BZ. \emph{Two}-dimensional IRs of C$_{6v}$ are represented by $\Gamma_7$, $\Gamma_8$ and $\Gamma_9$ which are constructed out of $\{\psi_3,\psi_5\}$, $\{\psi_1,\psi_4\}$, and $\{\psi_2,\psi_6\}$ eigen sub-space of C$_{6z}$ rotation operator respectively. $\Gamma_4$ and $\Gamma_5$($\Gamma_6$) are \emph{two} and \emph{one}-dimensional IRs of C$_{3v}$ along $k_z$-axis, constructed using $\{\psi_2,\psi_3\}$ and $\psi_1$ eigenstate of C$_{z}$.}
\label{symmcartoon}
\end{figure}

In this section, we explain the symmetry protection of Dirac and triply degenerate nodal point purely from group theoretical approach. 
{\par \it For C$_{6v}$ point group}: The ternary compound belonging to prototype BaAgBi possesses space inversion symmetry, six fold rotation (C$_6$) along the {\it k$_z$} axis and six mirror planes (3$\sigma_v$ \& 3$\sigma_d$). The $\Gamma$-A line lies in the $\sigma_v$ mirror plane and coincides with C$_6$ axis. $\sigma_d$ are those mirror reflection plane which bisects the angle between two $\sigma_v$ mirrors (see Fig.~\ref{symmcartoon}a). BaAgBi posses D$_{6h}$ point group symmetry at $\Gamma$ point, allows C$_{6v}$ little group along {\it k$_z$} axis in BZ. The C$_{6v}$ group contains the symmetry elements identity (E), sixfold, threefold and twofold rotation (C$_6$, C$_3$, and C$_2$) about {\it k$_z$~}axis. In the absence of SOC, the eigenvalues of C$_{6z}$ rotation operator are $e^{\frac{i2\pi}{6}n}$, where $n$ = 0 to 5 and the corresponding eigenstates are denoted
by $\psi_1$, $\psi_2$, $\psi_3$, $\psi_4$, $\psi_5$ and $\psi_6$ respectively. Under $y$-axis mirror ($a$ $\sigma_v$ mirror i.e, $xz$ plane in Fig.~\ref{symmcartoon}) transformation, $\psi_1$, and $\psi_4$ states remain same but $\psi_2$ changes to $\psi_6$ (i.e, $\tilde{\sigma}_v \psi_2 \rightarrow \psi_6$) and $\psi_3$ changes to $\psi_5$ (i.e, $\tilde{\sigma}_v \psi_3 \rightarrow \psi_5$). Here, tilde($\sim $) over the symmetry elements means the operation as opposed to just point group elements. Note that $\tilde{C}_{6}$ does not commute with $\tilde{\sigma}_v$ ($xz$ mirror plane in Fig.~\ref{symmcartoon}(a)). This \emph{non-commutation} of $\tilde{C}_{6}$ and $\tilde{\sigma}_v$ plays a crucial role here because if these two operators commute, then the operation of $\tilde{\sigma}_v$ on $\psi_2$ or $\psi_3$ can not convert these two states into $\psi_6$ or $\psi_5$ respectively (see Fig.~\ref{symmcartoon}(a)). Furthermore, the Hamiltonian ($H$) commutes with $\tilde{\sigma}_v$ i.e, $[H, \tilde{\sigma}_v]= 0$. This commutation gives-

\begin{subequations}\label{E2equalsE6}
\begin{align}
H \tilde{\sigma}_v \psi_2 - \tilde{\sigma}_v H \psi_2 & =  0 \\
H \psi_6  - E_2 \tilde{\sigma}_v\psi_2 & =  0  \\
E_6 \psi_6 - E_2 \psi_6 & =  0 
\end{align}
\end{subequations}

Hence, above algebra in equation~\eqref{E2equalsE6} implies that the energy eigenvalue, $E_2$ and $E_6$ corresponding to rotation eigen states $\psi_2$ and $\psi_6$ respectively, are the same. Similarly, it can be shown that $E_3 = E_5$, where $E_3$ and $E_5$ are the energy eigenvalue corresponding to $\psi_3$ and $\psi_5$ eigen states respectively. Thus, \emph{non-commutation} of $\tilde{C}_{6z}$ and $\tilde{\sigma}_v$ enforces the two double degenerate bands spanned by $\{\psi_2,\psi_6\}$ and $\{\psi_3,\psi_5\}$ along the {\it k$_z$} axis which is invariant under both $\sigma_v$ and C$_{6}$. However, $\psi_1$ and $\psi_4$ can not be coupled together following the above methodology. Hence they ($\psi_1$ \& $\psi_4$) remain non-degenerate. From this analysis, we can conclude that C$_{6v}$ symmetry group must contain both \emph{one} and \emph{two-}dimensional irreducible representations (IRs) when SOC is ignored. This is indeed true as shown in supplement Fig.~S1,\citep{supp} which is a band structure for BaAgBi in the absence of SOC.

{\par} From Figure~\ref{symmcartoon}(a), it is tempting to guess that $\psi_1$ and $\psi_4$ can be transformed to each others using another mirror plane $\sigma_d$ ($yz$-plane) and hence they can be degenerate as discussed for other states in the above texts. But, however, this is not true. This is because, these two states ($\psi_1$ and $\psi_4$) are eigen states of a C$_{2z}$ operation, and C$_{2z}$ operation commutes with $\sigma_d$ operation as pictorially shown in Fig.~\ref{symmcartoon}(a). This \emph{commutation} implies, $\psi_1$ and $\psi_4$ are eigenstate of $\sigma_d$ operator too. Therefore, the commutation refuses the transformation between $\psi_1$ and $\psi_4$ under the act of $\sigma_d$ ($yz$-plane) mirror and thus does not allow them to be degenerate. Interestingly, this commutation will play another important role when SOC will be considered which is going to be discussed in the following paragraph.
 
{\par} Now, in the presence of SOC, the eigenvalues for C$_{6z}$ rotation operator are $e^{\frac{i2\pi}{6}(n+1/2)}$ with $n$ = 0 to 5 and we label the corresponding eigen states in the same identical designation ($\psi_1$ to $\psi_6$) as before given for without SOC case. With SOC in consideration, the act of $\sigma_v$ operation follows a slightly modified transformation rule and it transforms $\psi_2$ and $\psi_3$ as; $\tilde{\sigma}_v \psi_2 \rightarrow i\psi_6$ and $\tilde{\sigma}_v \psi_3 \rightarrow i\psi_5$. But, however, using the same arguments as before in without SOC case, it can be shown that the earlier discussed \emph{non-commutation} of $\tilde{C}_{6z}$ and $\tilde{\sigma}_v$ still enforces the two double degenerate bands spanned by $\{\psi_2,\psi_6\}$  and  $\{\psi_3,\psi_5\}$ in presence of SOC along the {\it k$_z$} axis which is invariant under both $\sigma_v$ and C$_{6z}$ operation. 

{\par} Unlike without SOC case, here $\psi_1$ and $\psi_4$ make a degenerate eigen space in the presence of SOC. Since in earlier case we have discussed $[\tilde{C}_2,\tilde{\sigma}_d]=0$, we now define a new operator as the product of $\tilde{C}_2$ and $\tilde{\sigma}_d$ i.e, $\zeta$=$\tilde{C}_2\tilde{\sigma}_d$. Now, $\zeta^2=1$ due to the commutation of $\tilde{C}_2$ \& $\tilde{\sigma}_d$ and their square equals to -1 when SOC is considered. In the presence of TRS ($\mathcal{T}$), we now define a new operator, $\Theta=\zeta \mathcal{T}$. As $\mathcal{T}^2$ = -1 in the presence of SOC, we have $\Theta^2=\zeta^2 \mathcal{T}^2= -1$ along {\it k$_z$} axis which is invariant under both C$_2$ and $\sigma_d$. Thus, along {\it k$_{z}$} axis, $\Theta^2=-1$ acts as a local Kramer's theorem and enforce double degeneracy of $\psi_1$ and $\psi_4$ on the C$_{2z}$ axis. Therefore, C$_{6v}$ allows \emph{only} two-dimensional IRs along C$_6$ axis. Thus, any accidental band crossing on the C$_6$ axis will form a four fold degenerate Dirac node (as pictorially shown in Fig.~\ref{symmcartoon}(b) cartoon diagram). The Dirac nodes are protected from band hybridization induced gap opening if the two doubly degenerate bands belong to different irreducible representations of $\tilde{C}_{6v}$. Up to now we have only considered the C$_{6v}$, the little group of $\textbf{k}$ along $\Gamma$-A in our discussion. As we have discussed, C$_{6v}$ by itself forms a doubly degenerate eigen space along $\Gamma$-A when SOC is considered. Interestingly, our proposed materials have $P6_3/mmc$ space group possesses inversion symmetry ($\mathcal{I}$) too. Now, the act of Kramer's theorem says that $\mathcal{I}$ and TRS together enforce all the bands to become doubly degenerate over the whole BZ including $\Gamma$-A direction. However, our analysis for C$_{6v}$ subgroup is more general and it shows C$_{6v}$ could alone give Dirac like band crossings despite of having inversion symmetry. Thus our analysis is more general and applicable for any system which has C$_{6v}$ point group.

\begin{figure}[t!]
\centering
\includegraphics[width=\linewidth]{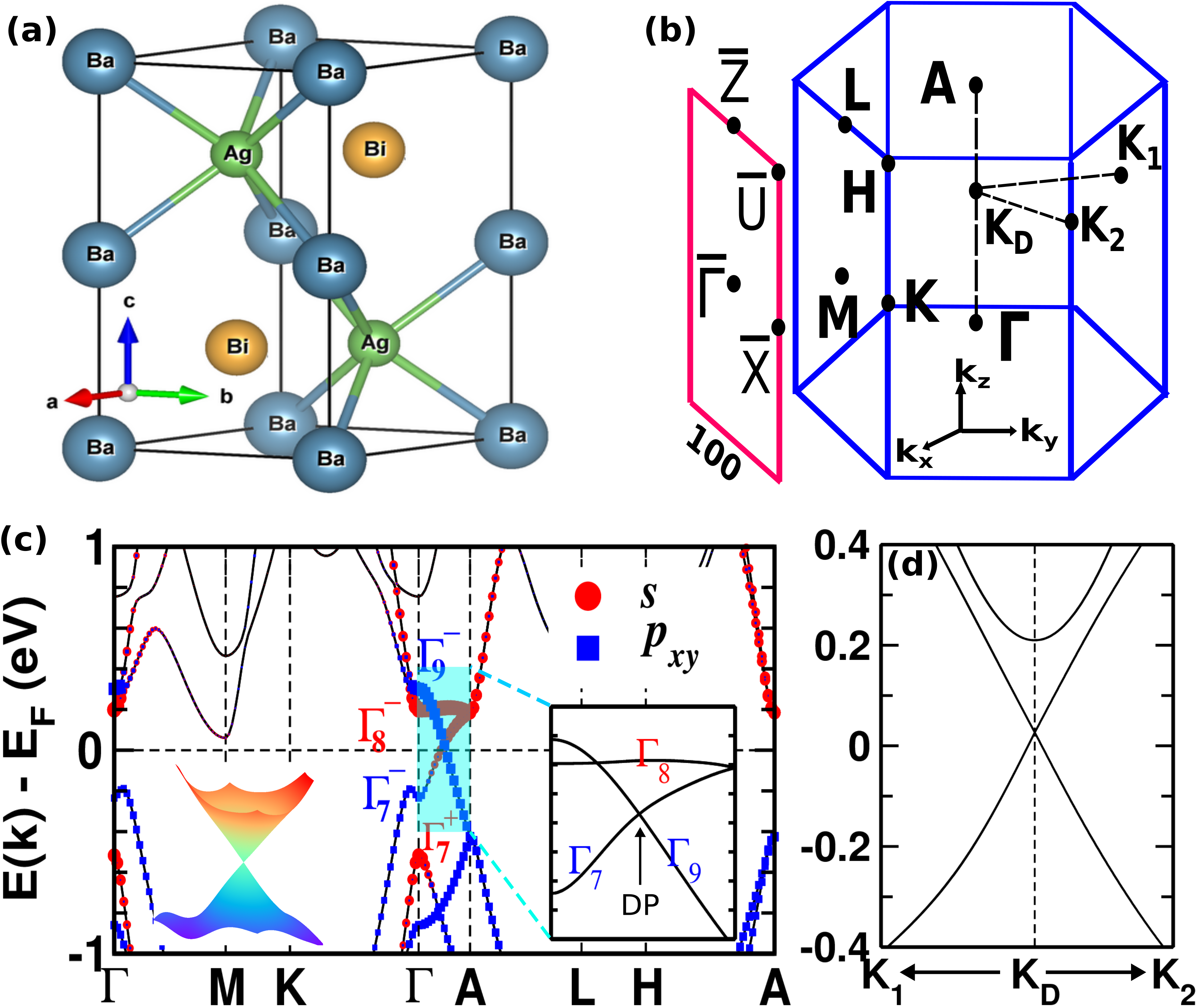}
\caption{(Color online) For BaAgBi (a) crystal structure (space group P6$_3$/mmc), (b) Bulk Brillouin zone (BZ) and (100) Surface BZ with high symmetry points. (c) Bulk band structure within HSE06$+$spin-orbit coupling along full K-path. The inset in (c) are enlarged view of shaded area. Three dimensional view of Dirac cone is also shown in inset of (c). $\Gamma_i,s$ are the irreducible representations of band characters. (d) Band structure around Dirac point (DP), in the \textit{k$_x$-k$_y$} plane, where (0,0,$\pm$0.239$\frac{2\pi}{c}$).  }
\label{fig2}
\end{figure}

{ \par \it For C$_{3v}$ point group}: In the above, we have discussed about the dimensional degeneracy of IRs for a C$_{6v}$ point group symmetry (along \emph{k$_z$}-axis) which is of course a subgroup of D$_{6h}$. Now, breaking of inversion symmetry elements and three $\sigma_d$ mirror planes in D$_{6h}$ point group symmetry, transforms it into a D$_{3h}$ point group symmetry. Hence, for a D$_{3h}$ point group symmetry based system, the little group of \textbf{k} on the k$_z$ axis is now C$_{3v}$ instead of C$_{6v}$. The symmetry elements for C$_{3v}$ point group are identity (E), three-fold rotation (C$_3$) and three $\sigma_v$ mirror. The combination of a mirror plane and a C$_{3z}$ axis is schematically shown in Fig.~\ref{symmcartoon}(c). Similar to the above analysis for C$_{6v}$, one can also obtain the band degeneracy along a C$_3$ axis for C$_{3v}$ symmetry, will be discussed in the followings. Without SOC, the eigenvalues of $\tilde{C}_{3}$ are 1, $e^{i\frac{2\pi}{3}}$ and $e^{-i\frac{2\pi}{3}}$. Consider $\psi_1$, $\psi_2$ and $\psi_3$ are the eigen states of $\tilde{C}_{3}$ corresponding to the eigenvalues $1$, $e^{i\frac{2\pi}{3}}$ and $e^{-i\frac{2\pi}{3}}$ respectively. Now, $\tilde{\sigma}_v$ ($xz$ mirror plane) transforms $\psi_2$ to $\psi_3$ (i.e., $\tilde{\sigma}_v \psi_2 \rightarrow  \psi_3$). Thus $\tilde{\sigma}_v$ and $\tilde{C}_{3z}$ do not commute in the subspace of $\psi_2$ and $\psi_3$ which enforces double degenerate bands along C$_3$ axis, spanned by \{$\psi_2,\psi_3$\} states. However, $\tilde{\sigma}_v$ ($xz$ mirror plane) left $\psi_1$ unchanged and it appears as a singly degenerate band along C$_3$ axis. Thus for C$_{3v}$ point group, both \emph{two} and \emph{one} dimensional IRs are possible along the C$_3$ axis. Further, inclusion of SOC only changes the eigen values of $\tilde{C_3}$ (-1, $e^{i\frac{\pi}{3}}$ and $e^{-i\frac{\pi}{3}}$) and doubled the number of bands. The IRs still remain in both \emph{one} and \emph{two}-dimensional in double group of C$_{3v}$. Thus accidental crossing of singly degenerate band (formed by $\psi_1$) and doubly degenerate band (formed by $\psi_{2}$ and $\psi_{3}$) allows symmetry protected three fold degeneracy along C$_3$ axis, as schematically shown in Fig.~\ref{symmcartoon}(d). A similar analysis for $C_{3v}$ has earlier been drawn in Ref.[\onlinecite{nexus}] to describe the triple point nexus fermion.

\begin{figure}[t]
\centering
\includegraphics[width=\linewidth]{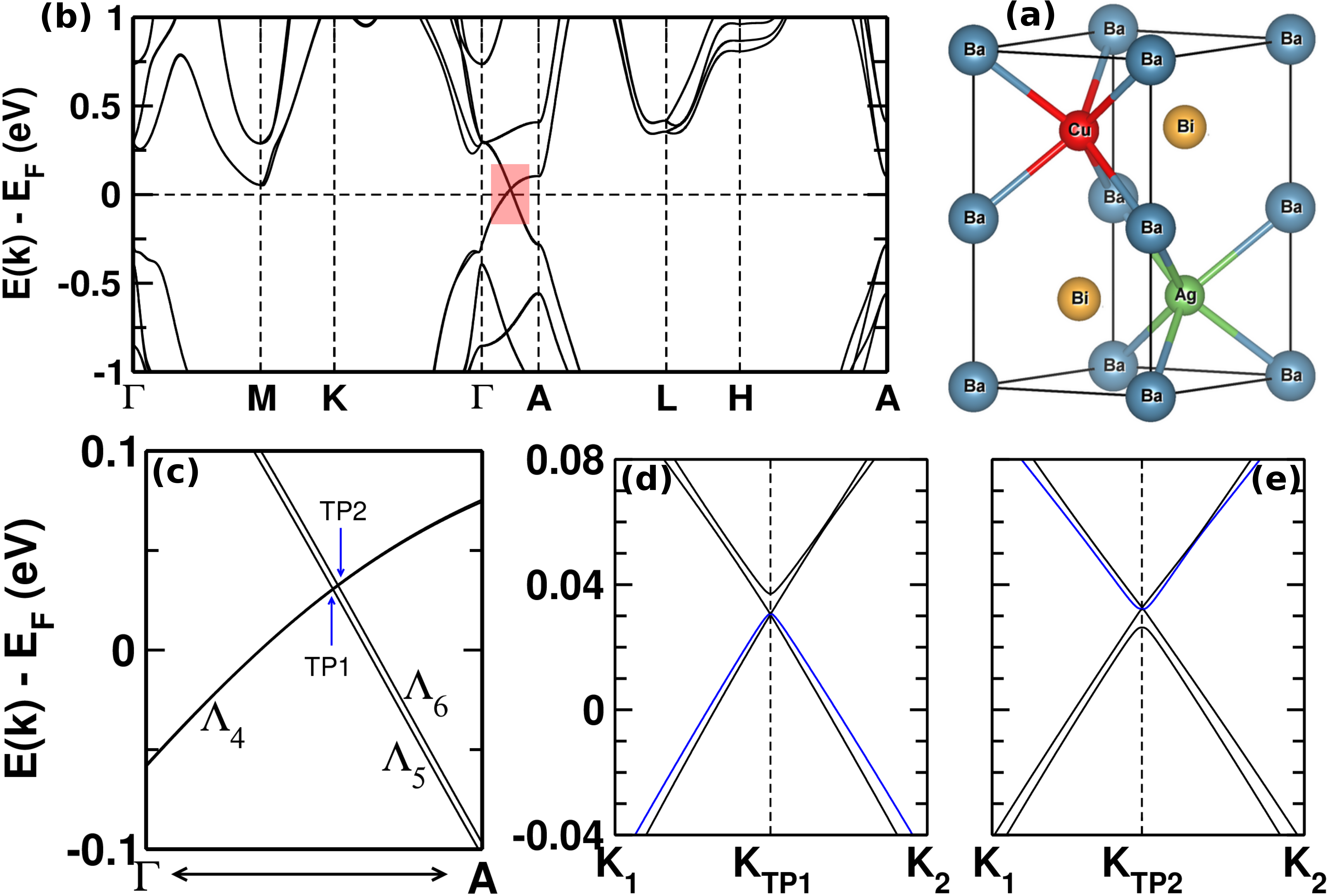}
\caption{(Color online) For BaAg$_{0.5}$Cu$_{0.5}$Bi. (a)Crystal structure with D$_{3h}$ point group symmetry. (b) Band structure along full K-path (a). The shaded box area in (b) is shown in (c) along with irreducible representations $\Lambda_i^{'s}$ of bands. (d,e) Band structure along around the two triply degenerate nodal points (TP1 and TP2) in the \textit{k$_x$-k$_y$} plane. }
\label{fig3}
\end{figure}

\section{Results and Discussions}
\subsection{Dirac semimetal (DSM)}
Considering BaAgBi as representative system, we have done band structure calculation within HSE06 hybrid functional. Results for the other compounds are presented in supplement (SM).\cite{supp} Figure~\ref{fig2}(a) shows crystal structure of BaAgBi (space group P6$_3$/mmc). Barium (Ba), silver (Ag) and Bismuth (Bi) sit at Wyckoff positions 2a, 2c and 2d respectively in the unit cell. Other ternary compounds presented in SM\cite{supp} crystallize in the same structure with space group P6$_3$/mmc. The bulk and (100) surface Brillouin zone (BZ) with the high symmetry points are shown in Fig.~\ref{fig2}(b). Figure~\ref{fig2}(c,d) shows the electronic band structure for BaAgBi calculated using the HSE06 functional and spin-orbit coupling (SOC). The irreducible representations (IRs) of the bands are depicted by $\Gamma_i$'s. $\Gamma_{i}^{\pm}$ represents IRs with even(odd) parity states. At $\Gamma$ point the point group symmetry is D$_{6h}$ and all the bands are two dimensional IRs of D$_{6h}$ subgroup. \AAEDITOKAY{A pair of bands ($\Gamma_{7}^{+}$ and $\Gamma_{8}^{-}$) at $\Gamma$ point are of s-orbital like orbital character}{At this point, $\Gamma_{7}^{+}$ and $\Gamma_{8}^{-}$ bands possess {\it s-}like orbital character.} Whereas the other two bands, $\Gamma_{7}^{-}$ and $\Gamma_{9}^{-}$ possess {\it p$_x$-p$_y$} like orbital character.

{\par}The point group symmetry along $\Gamma$-A line is C$_{6v}$. Hence, as discussed earlier, each band along the C$_6$ axis ($\Gamma$-A) has doubly degenerate IRs $\Gamma_i$ of C$_{6v}$ subgroup. The band linking rule suggests $\Gamma_{i}^{\pm}$ states at $\Gamma$ point transformed as $\Gamma_{i}$ (IRs of C$_{6v}$ subgroup) along C$_6$ axes. The inter-crossing of $\Gamma_{7}$ and $\Gamma_{9}$ band along C$_6$ axis gives rise to a Dirac nodal point (DP) very close to the Fermi level (E$_F$), as shown in Fig.~\ref{fig2}(c). Interestingly there is one more Dirac node which forms due to the crossing of $\Gamma_{8}$ and $\Gamma_{9}$ IRs near the $\Gamma$ point at around 0.2 eV above the E$_F$. The surface signature of this Dirac node, however is not trackable as it is significantly buried within the bulk density. The dispersion around the Dirac node DP, near E$_F$ in k$_x$-k$_y$ plane is shown in Fig.~\ref{fig2}(d) along K$_1$-K$_D$-K$_2$ direction, where K$_D$=(0,0,$\pm$0.239$\frac{2\pi}{c}$). The three dimensional view of the Dirac cone is also shown in Fig.~\ref{fig2}(c) inset. Since, these ternary compounds host space inversion symmetry ($\mathcal{I}$), all the band throughout the BZ are doubly degenerate Kramer's pair. Hence, the four-fold Dirac crossings are ensured by both C$_{6v}$ and $\mathcal{I}$. Notably, even in the presence of $\mathcal{I}$ breaking perturbation, the C$_{6v}$ symmetry alone can protect the Dirac nodes. Such scenario can also be observed in noncentrosymmetric materials belonging to P6$_3$mc (No. 186) space group.

\begin{figure}[t]
\centering
\includegraphics[width=\linewidth]{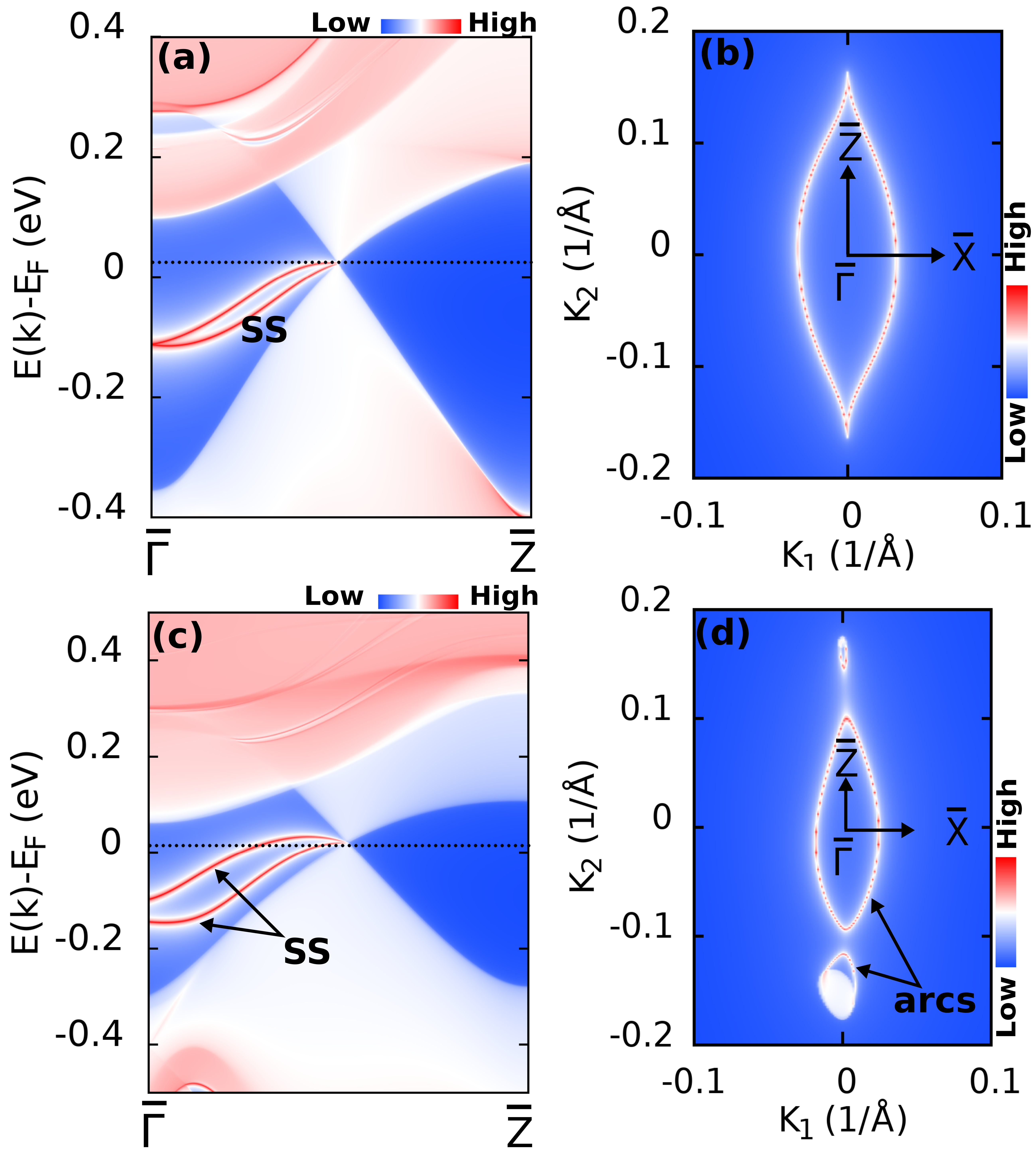}
\caption{(Color online) For BaAgBi (a) surface state (SS) on (100) miller plane emerging from Dirac points and (b) corresponding Fermi arcs. For BaAg$_{0.5}$Cu$_{0.5}$Bi (c) two distinct surface states (SS) appearing on (100) miller plane from two TPs and (d) corresponding Fermi arcs. Dotted line in (a,c) are the energy cut at which Fermi arcs are calculated. }
\label{fig4}
\end{figure}

\begin{figure}[htb!]
\centering
\includegraphics[width=\linewidth]{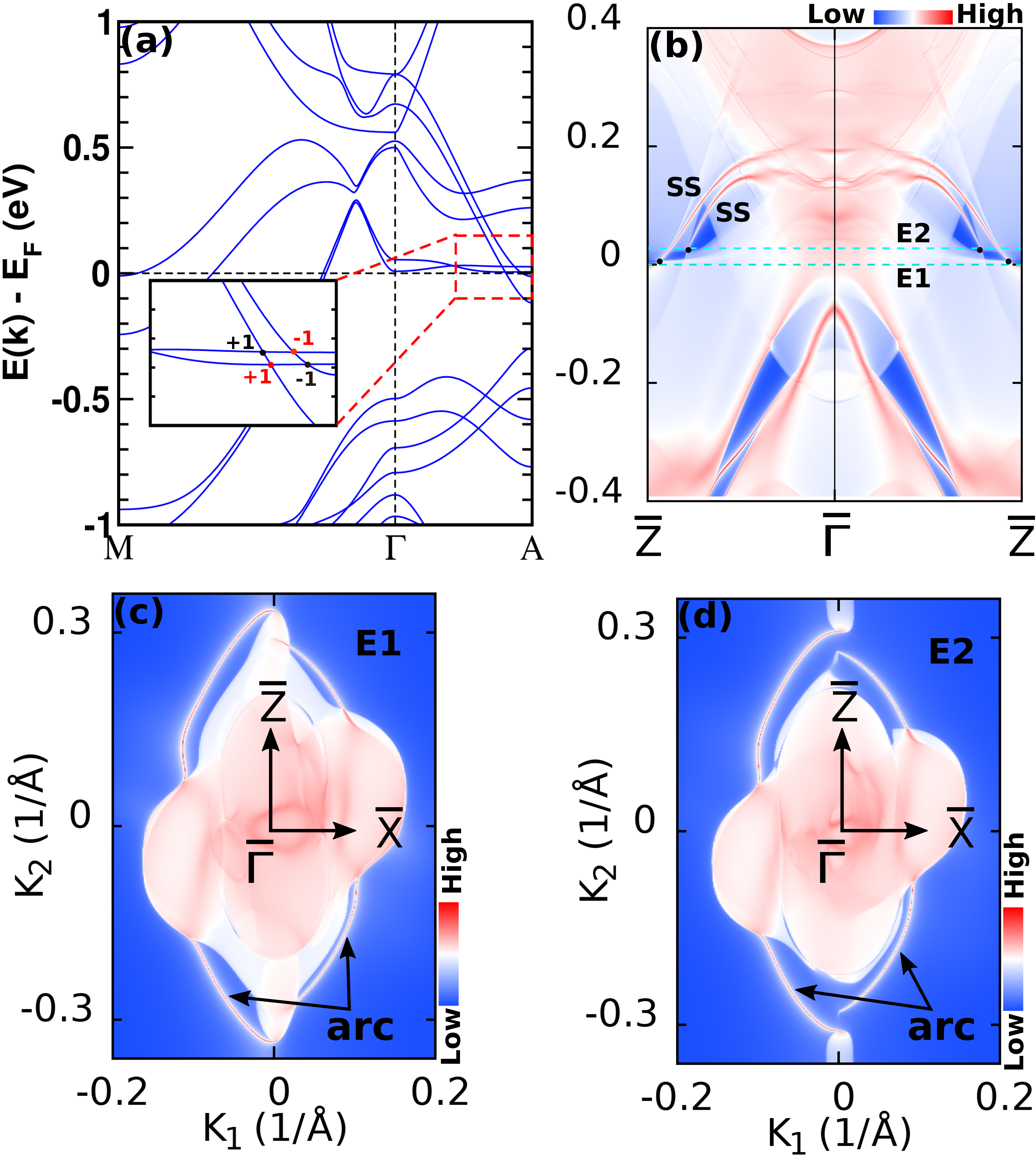}
\caption{(Color online) For Ba$_{0.5}$Eu$_{0.5}$AgBi. (a) Bulk band structure along M-$\Gamma$-A. It shows four Weyl nodes located in the vicinity of E$_F$ (shown inside red box). The inset in (a) shows a closer look of these Weyl node with their chirality. (b) (100) surface spectrum of Weyl phase and (c,d) corresponding Fermi arcs at two energy cuts E1 and E2 (as shown in (b)).   }
\label{fig5}
\end{figure}

\subsection{Triple point semimetal (TPSM)}
Our next objective is to explore if we can achieve other relevant topological phases, such as TPSM, WSM in a related compound with closely related chemical modification using the idea of symmetry breaking phase transformation mechanism. Here we discuss the transformation of DSM phase into TPSM phase. As discussed earlier, the C$_{3v}$ point group is an appropriate symmetry for giving three fold band degeneracy and hence we aim to introduce C$_{3v}$ in the crystal. Note that C$_{3v}$ is subgroup of D$_{3h}$ point group symmetry. To induce C$_{3v}$ point group symmetry along the {\it k$_z$} axis in the BZ, we substitute one out of two equivalent silver atom by a copper atom and hence form a mixed compound BaAg$_{0.5}$Cu$_{0.5}$Bi. The ordered crystal of structure of BaAg$_{0.5}$Cu$_{0.5}$Bi is shown in Fig.~\ref{fig3}(a). This structure indeed breaks the space inversion symmetry, transforms the six-fold (C$_6$) rotational symmetry to three-fold (C$_3$) and six mirror plane of symmetry to three mirror plane. Thus, the D$_{6h}$ point group symmetry for parent BaAgBi transforms to D$_{3h}$ for the mixed compound, BaAg$_{0.5}$Cu$_{0.5}$Bi and hence possesses the desired C$_{3v}$ subgroup along {\it k$_z$} axis. As expected, the band structure for BaAg$_{0.5}$Cu$_{0.5}$Bi indeed shows three fold degenerate band crossings along $\Gamma$-A direction as evident from Fig.~\ref{fig3}. Along {\it k$_z$} axis the IRs for C$_{3v}$ little group are of two dimensional $\Lambda_4$ and one dimensional $\Lambda_{5,6}$. Due to the Copper substitution in parent BaAgBi, the $\Gamma_7$ and $\Gamma_9$ IRs along $\Gamma$-A transform as $\Lambda_4$ and $\Lambda_{5,6}$ in the \textcolor{black}{mixed compound} case. Therefore, the inter-crossings of IRs along $\Gamma$-A direction give rise to a pair of triply degenerate nodal point (TDNP), namely TP1 and TP2 as shown in Fig.~\ref{fig3}(c). Furthermore, Fig.~\ref{fig3}(d,e) shows the band structure around each TP along a certain direction in {\it k$_x$-k$_y$} plane, which further confirms that the three fold crossing are formed out of three distinct non-degenerate bands. It is important to note that, unlike other TPSM the location of TDNP in the present case is very close to E$_F$ with no extra trivial Fermi pockets. This makes this compound a superior candidate for possible experimental investigations in future. 

\textcolor{black}{It is important to note that we have considered an ordered structure for BaAg$_{0.5}$Cu$_{0.5}$Bi where Cu is allowed to only occupy a specific Wyckoff position in the parent structure. However, there is a possibility for the random disorder between Ag and Cu as well. In order to check this, we have simulated  the energetics of both ordered (as chosen above) as well as random disordered configuration using Coherent Potential Approximation (CPA)\cite{CPA} of Ag$_{0.5}$Cu$_{0.5}$ in BaAg$_{0.5}$Cu$_{0.5}$Bi alloy. The ordered configuration is found to be energetically more stable, and hence our choice for the performed simulation.}
 
\subsection{Surface excitations for DSM \& TPSM}
{\par} Topological semimetals are known to show exciting topological signatures on their surfaces originated via their non-trivial band topology. In Fig.~\ref{fig4}, we showcase the surface spectra and Fermi arcs topology of the (100) side surface for DSM and TPSM phases. The surface BZ along with high symmetry points are depicted in Fig.~\ref{fig2}(b). Dirac points and three fold nodal points are projected at distinct positions along $\bar{\Gamma}-\bar{Z}$ line on the surface BZ. As obvious from Fig.~\ref{fig4}(a) for the DSM case, a two-fold degenerate surface states (SS) are seen to emerge from the Dirac node which lies near E$_F$. Since, the (010) surface lacks the space inversion symmetry, the degeneracy of the SS is lifted except at the time reversal invariant momenta (TRIM) $\bar{\Gamma}$ point. Dirac nodes are superposition of two Weyl nodes of opposite chirality and it is stabilized by the specific crystalline symmetries.\cite{Ashvin2018} Therefore, it is expected to observe a two-fold degenerate surface states from a Dirac node. Thus, a pair of close Fermi arcs appear from a Dirac node and nested between bulk projected Dirac points on the surface as shown in Fig.~\ref{fig4}(b). On the other hand for TPSM, the Dirac point splits into two TDNP along {\it k$_z$} axis and a non degenerate surface states appears from each TDNP as shown in Fig.~\ref{fig4}(c). The Fermi arc topology for the corresponding surface states is shown in Fig.~\ref{fig4}(d) at an isoenergy level shown by the dotted line in Fig.~\ref{fig4}(c). The Fermi arcs originated from two TDNPs are gaped along {\it k$_z$} axis. A branch of Fermi arc appeared from a TDNP and propagates through the surface BZ to connect the bulk projected surface TDNP. Another piece of the Fermi arc appears and merges into the bulk bands. A similar pattern of Fermi arc topology for  TPSM has previously been reported in Ref. \onlinecite{NaCu3Te2},\onlinecite{MgTaNbN}.

\subsection{Weyl semimetal (WSM)}
{\par} Breaking of either $\mathcal{I}$ or TRS (in our case), can lead to the emergence of WSM phase.\cite{PTbreaking1,PTbreaking2} In order to achieve this we dope parent BaAgBi with a magnetic element Eu and form an \textcolor{black}{mixed compound} Ba$_{0.5}$Eu$_{0.5}$AgBi. \textcolor{black}{ Note that, here we considered the ordered arrangement of Eu on the Ba site to get the Weyl phase. Even if one considers the disordered configuration, it does not matter much as long as the disorder does not create huge chemical imbalance that removes the Weyl nodes from the momentum space. This is because the Weyl nodes do not require any symmetry to be maintained}. Ferromagnetic order is considered in the \textcolor{black}{ mixed compound}. The ferromagnetism of Eu atom breaks the TRS while keeping the space inversion symmetry intact in the \textcolor{black}{ mixed compound}. To ensure the proper inclusion of {\it f} electron localization we perform PBE+U calculations\cite{HubbardU} with a Hubbard U$_{eff}$ (U=5 eV; J=1 eV) introduced in a screened Hartree-Fock manner. Our calculation shows a magnetic moment of 6.83 Bohr magneton ($\mu_b$) from the Eu-{\it f} orbital. As expected, due to the breaking of TRS, the four-fold Dirac node along the $\Gamma$-A direction split into a pair of two-fold Weyl nodes as shown in Fig.~\ref{fig5}(a). Band structure with few more choices of U$_{eff}$ (U-J=2, 3, 5, and 6 eV) values are shown in the supplement sec. IV.\cite{supp} All these plots shows the ensurance of such WSM along $\Gamma$-A direction , confirming the robustness of Weyl nodes. Weyl nodes, act as source or sink of Berry curvature, always come in pairs with opposite chiral charge obeying the Nielsen-Ninomiya theorem.\cite{Nielsen} We calculate the Chern number or chirality associated with these Weyl nodes as shown in the inset of Fig.~\ref{fig5}(a). The black(red) dots represents the Weyl nodes of opposite chirality $\pm$1. It is well known and expected that Weyl nodes are connected by open Fermi arcs. To observe this, we projected these nodes of Ba$_{0.5}$Eu$_{0.5}$AgBi on (100) surface where the Weyl nodes of opposite chirality falls at different locations. In contrast, when projected onto the (001) hexagonal surface of Ba$_{0.5}$Eu$_{0.5}$AgBi, these Weyl nodes of opposite chirality fall on each other, leading to the disappearance of Fermi arcs. Figure~\ref{fig5}(b) shows the (100) surface spectrum representing a pair of states (SS) emerging from two Weyl nodes of opposite chirality (denoted by black color dots both in Fig.~\ref{fig5}(a)~\&~\ref{fig5}(b)). However, surface signature from other pair of Weyl nodes (denoted by red color dots in Fig.~\ref{fig5}(a) inset) are not identified here, because they are probably dispersed within the bulk of the spectrum. The Fermi arcs topology at two different energy scale (represented by dashed line in Fig.~\ref{fig5}(b)) are shown in Fig.~\ref{fig5}(c,d). As discussed above and unlike DSM, a pair of open Fermi arcs appeared from two Weyl nodes, which is one of the striking feature of WSM. These open Fermi arcs stick at two singular points where the surface projection of bulk Weyl nodes occur. Unlike DSM, the Weyl nodes and open Fermi arcs in WSMs are protected by non-trivial non-zero Chern number. 

\section{Effective model Hamiltonian}
In order to better understand the emergence of DSM phase, we now discuss the band crossings using a \textbf{k.p} Hamiltonian derived using the method of invariants (similar to those used in Na$_3$Bi\cite{A3Bi2012} and Cd$_3$As$_2$\cite{Cd3As22013}). Since, the low energy states around the Dirac node are governed by {\it s-} and {\it p-}like atomic orbitals, a 4$\times$4 Hamiltonian for D$_{6h}$ point group is set up using $|S^+_\frac{1}{2},\frac{1}{2}\rangle$, $|P^-_\frac{3}{2},\frac{3}{2}\rangle$, $|S^+_\frac{1}{2},-\frac{1}{2}\rangle$, $|P^-_\frac{3}{2},-\frac{3}{2}\rangle$ as basis set. The superscript $\pm$ in the basis set represents the parity of the states.

\begin{equation}\label{Hk}
H(\textbf{k}) = \epsilon_{0}(\textbf{k})\mathbb{1} + \begin{pmatrix} 
M(\bf k) & Ak_{+} & Dk_{-} & 0 \\
Ak_{-} & -M(\bf k) &0 & 0  \\
Dk_{+} & 0 & M(\bf k) &  Ak_{-}   \\
0 & 0 & Ak_{+} & - M(\bf k)  \end{pmatrix}
\end{equation}

Here, $ \epsilon_{0}(\textbf{k}) = C_0 + C_1k^{2}_{z} + C_2(k^{2}_{x}+k^{2}_{y})$, $A(\textbf{k}) = A_0 + A_1k^{2}_{z} + A_2(k^{2}_{x}+k^{2}_{y})$, $  M(\textbf{k}) = -M_0 + M_{1}k^{2}_{z} + M_2(k^{2}_{x}+k^{2}_{y})$, $k_{\pm} = k_{x} \pm ik_{y}$. To ensure the inverted band order $M_0$, $M_1$, and  $M_2$ takes the value greater than {\it zero}. The finite value of parameter $D$ ensures the absence of space inversion symmetry. For centro-symmetric system, $D=$0. The eigenvalues of $H(\textbf{k})$ for centro-symmetric system has the following form,

\begin{eqnarray}\label{eigenval}
E(\textbf{k}) = \epsilon_{0}(\textbf{k}) \pm \sqrt{M(\textbf{k})^2 + A^2k_+k_-} 
\end{eqnarray}

The eigenvalue equation in \eqref{eigenval} gives gapless solution at $\textbf{k}_d = (0,0,k_z=\pm\sqrt{M_0/M_1})$, which are the positions of Dirac nodes on $\textit{k}_z$ axis. By fitting the  eigenvalue equation \eqref{eigenval} in the vicinity of Dirac node of our first-principle band structure of BaAgBi, we get, $C_0=0.036~$eV , $C_1=-0.204~$eV $\AA^2$, $C_2=0.238~$eV $\AA^2$, $M_2=82.40~$eV $\AA^2$, $A=4.798~$eV $\AA$. A chosen value of $M_1 = 10.0~$eV $\AA^2$ gives $M_0=0.252~$eV using the condition $M_0/M_1 = (0.239\frac{2\pi}{c})^2 \AA^{-2}$, where $c=9.452 \AA$.

{\par} We can further apply a \textit{Zeeman} field ($h \sigma_z \otimes \tau_z$, where $h$ is field strength) in the Dirac Hamiltonian $H(\textbf{k})$ to break the time reversal symmetry ($\mathcal{T}$). In that case, the Hamiltonian $H(\textbf{k})$ takes the form; 


\begin{equation}\label{zeeman}
\begin{aligned}
H_{mag}(\textbf{k}) =& H(\textbf{k}) + h \sigma_z \otimes \tau_z \\
\end{aligned}
\end{equation}

The upper (lower) block of $H_{mag}$ forms a $2\times2$ Weyl Hamiltonian of opposite Chern number. Such splitting of Dirac node under \textit{Zeeman} field gives rise to Weyl nodes on the \textit{$k_z$} axis. \AAEDITOKAY{This is another way of realizing Weyl phase from Dirac semimetal instead of breaking $\mathcal{T}$ via doping with magnetic elements as shown in earlier section.}{}

\section{Conclusion} 
In summary, using accurate hybrid functional calculations and symmetry analysis, we revisited the existence of Dirac like band crossings in few broadly experimentally synthesized materials belonging to the space group $P6_3/mmc$. Our DFT calculations reveal that the use of relatively more accurate hybrid functional is extremely crucial and important in order to correctly search promising topological quantum materials with unavoidable band crossings. Less accurate exchange correlation functionals are shown to be inadequate to properly predict the band order. Further, by alloy engineering mechanism we show that lattice symmetry breaking can help to achieve various distinct topological phases in a single compound. BaAgBi, as a representative candidate, is predicted to be an ideal Dirac semimetal (DSM) with the Dirac node located in the vicinity of the E$_F$ and no extra trivial Fermi pockets. Further, Copper and Europium doping in BaAgBi gives the birth of triple point and Weyl semimetallic phases with the nodal points again located  at/near the Fermi level. We have also studied bulk-boundary correspondence on the surface of these symmetry protected topological phases. $\textbf{k.p}$ model Hamiltonian is further utilized to better understand the symmetry protected band degeneracy in BaAgBi. In conclusion, we strongly believe that while symmetry of the crystal structure is essential to protect the band crossings, more accurate exchange-correlation functional in any DFT calculation is an important necessity to determine the correct band order before judging a material as topologically non-trivial and hence for possible future experiments.

\section*{Acknowledgement}
 CKB acknowledges IIT Bombay for financial support in the form of teaching assistantship. CM acknowledge MHRD-India for financial support.  AA acknowledge DST-SERB (Grant No. CRG/2019/002050) for funding to support this research. BP acknowledge DST SERB-India. We thank IIT Bombay for high performance computing facilities.




\end{document}